\documentstyle[aps,epsf]{revtex}

\begin{document}
\title{Decoherence effects of motion-induced radiation}

\author{Paulo A. Maia Neto $^{1}$ \thanks{e-mail: \tt pamn@if.ufrj.br}
and Diego A.\ R.\ Dalvit $^{2}$  \thanks{e-mail: \tt dalvit@lanl.gov}}

\address{$^1$ \it Instituto de F\'{\i}sica, UFRJ, Caixa Postal 68528, 
21945-970 Rio de Janeiro, Brazil \\
$^2$ \it T-6, Theoretical Division, MS B288, Los Alamos National
Laboratory, Los Alamos, NM 87545}

\maketitle

\begin{abstract}
The radiation pressure coupling with vacuum fluctuations gives rise to
energy damping and decoherence of an oscillating particle. 
Both effects result from the emission 
of pairs of photons, a quantum effect related to
the fluctuations of the Casimir force. We
discuss different alternative methods for the computation
of the decoherence time scale.
We take the example of a spherical
perfectly-reflecting particle, and consider the zero and high temperature
limits. We also present short general reviews on decoherence and dynamical Casimir effect.
\end{abstract}

\section{Introduction and brief summary of decoherence theory}

The understanding of the quantum-to-classical transition has been the
subject of extensive research\cite{giulini,WHZ1}. The core of the problem
is that the Hilbert space of allowed states of a quantum system is huge, 
whereas the set of states with associated classical properties is a tiny subset
of the whole Hilbert space. Some questions that naturally arise are the 
following: which mechanism is responsible for the classical appearance of
macroscopic and mesoscopic quantum systems? How are those few classical 
states selected from the huge Hilbert space?
The common wisdom is that classicality is
an emergent property induced on subsystems by their environment. 

The interaction between a system S and its environment E creates entanglement
(i.e. non reducible correlations) between the states of the system and those
of the environment. 
Imagine that at a given time (say $t=0$) the state $|\Psi(t)\rangle$ of S+E is a product state, 
that is, there are no initial correlations. 
We have $|\Psi(t=0)\rangle = |s\rangle |\epsilon\rangle$, where the 
first ket corresponds to the system state (assumed for simplicity to be in a 
pure state), and the second one to the environmental state, also assumed pure.
When the two parts begin to interact, entanglement is generally produced. This
means that at a later time $t$, the state $|\Psi(t)\rangle$ will be given by 
a linear superposition of the form

\begin{equation}
| \Psi(t) \rangle = |s_1\rangle |\epsilon_1\rangle +
                    |s_2\rangle |\epsilon_2\rangle + \ldots\label{entangled}
\end{equation}
where $\{|s_i\rangle\}$ and 
$\{|\epsilon_i\rangle\}$ are states of the system and 
environment Hilbert spaces, respectively. 
If the interaction is such that the states $|\epsilon_i\rangle$ 
become approximately orthogonal 
($\langle \epsilon_n|\epsilon_m\rangle \approx \delta_{nm}$), then
interference between the system states $|s_n\rangle$ and $|s_m\rangle$
will not be observed. These set of states $\{ |s_i\rangle\}$ usually have 
classical properties.
Any quantum superposition of them is a  
non-classical state, and
quickly decays away into a statistical mixture of the states.
The coherence of the phase relation between the components of the superposition
is lost, and this process is accordingly known as {\it decoherence}.
In other words, the environment monitors the different classical
alternatives for the system (the different states $|s_i\rangle$), thereby
providing which-way information, even though such information is usually
unaccessible to the observer.
The set of states $\{ |s_i\rangle\}$ 
are called {\it pointer states} \cite{WHZ2}, 
and they are the states
within the huge Hilbert space of the system that become less 
entangled with the environment. Perfect pointer states are those that produce
no entanglement at all, so that an initial product state of S+E will remain
a product state throughout the interaction time, which means that those states
are robust and stay unperturbed by the interaction. All this 
will be illustrated in Section~3 in the particular
case where the environment is the radiation field at zero temperature 
(vacuum field) and radiation pressure is responsible for the coupling
between system (a mirror) and the environment.

A possible method to identify  pointer states is called the
`predictability sieve criterion' \cite{WHZ3}, 
which is based on the fact that pointer 
states are the ones that produce least entropy and remain most pure. Let us 
explain these concepts. 
The evolution of the closed combined system S+E is unitary, so that the purity
of the whole state $| \Psi\rangle$ is preserved, i.e.
$P(\rho_{S+E})={\rm  Tr} \rho_{S+E}^2 =1$ for all times. 
However, the purity of S is not preserved. 
To show it one needs to calculate the reduced density matrix of the subsystem
S by tracing out the environmental degrees of freedom, 
$\rho_S = {\rm Tr}_E \rho_{S+E}$, and then $P(\rho_S)<1$. The loss of purity
can be associated with a loss of information about the system state. 
When no measurement involving the environment is made, 
the density matrix $\rho_S$ 
contains the state of knowledge of an observer about the
system, and purity is a measure of that knowledge. Initially,
there is full knowledge of the system state, which is described by a single
ket state. Subsequently, the interaction with the environment produces 
entanglement, and part of the information about the system S is lost to the
environment, causing a decrease in the purity of the system.
One can also measure the information content of $\rho_S$ 
through the von Neumann
entropy, $S(\rho_S)=-{\rm Tr} \rho_S \log \rho_S$. Initially one has full 
information, and entropy is identically zero; as time goes on, information
is lost and entropy is produced. Pointer states are least affected by the 
environment, so their information
content is preserved and hence they produce the least entropy. 
The idea is then to take every 
state in the
Hilbert space of S, calculate the von Neumann entropy at time $t$ produced 
via interaction with the environment starting from the given state of the 
system, and order the initial states in a tower of increasing entropy. The
states that lie at the bottom of such a tower are candidates for pointer
states. Finally, one must check that those states remain at the bottom when
the time $t$ when entropy is calculated is changed, so that the states are
robust pointers. If one can satisfy these conditions, one gets the
pointer states of the system S. If not, there are no pointer states for the 
problem. When these ideas are put into practice, it is more convenient to work
with the linear entropy, defined as 
$S(\rho_S)=1-P(\rho_S)=1-{\rm Tr} \rho_S^2$.

The determination of pointer states for a given problem depends both
on the system and environment dynamics, as well as on their interaction.
There are three different regimes: 1) When the system Hamiltonian is 
irrelevant, pointer states are given by the eigenstates of the system's
operators contained in the interaction Hamiltonian. A typical example
is that of a measuring apparatus (the system) that has no internal dynamics,
measuring an external reservoir (the environment), say a photocurrent. 
2) When the system dynamics as well as the interaction are relevant, pointers 
come from an interplay between the two. The most thoroughly studied example
is that of quantum Brownian motion (QBM), in which a particle is coupled to a 
set of harmonic oscillators in a thermal state, thereby suffering decoherence
\cite{QBM}.
Although the coupling between the system and the environment is of the type
position-position, pointer states are not position eigenstates of the system
because the self dynamics of the harmonic oscillator interchanges position
and momentum every quarter of a cycle. It turns out that the interplay
between interaction and self dynamics leads to coherent states as pointers \cite{ZHP}. 
This second case is also the relevant one for this paper.
3) Finally, when the environment evolves much slower
than the system, pointers may correspond to energy eigenstates of the system's
Hamiltonian \cite{ZP}.

In the above we have ignored the information contained in the environmental
state, and that is the reason why one traces over the environmental degrees
of freedom in order to find the reduced density matrix of the system.
However, the information lost to the environment could be, in principle,
intercepted and recovered. Performing measurements on the environment one 
may extract information about the decohering system. In \cite{CME} it is
shown that the preferred pointer states remain unchanged, even when that
information is kept and modifies the dynamical
evolution of the system. 

Another related way to study the dynamical process underlying in the
quantum-to-classical transition is via phase space representations of the
reduced quantum dynamics for the system S. Among the many possible 
representations, one of particular interest is the Wigner function 
$W(x,p)$, which is defined as a Fourier transform of the reduced density matrix.
$W(x,p)$ is a pseudo probability 
distribution in phase space, and encapsulates the quantum coherence of the system in
interference fringes that take both positive and negative values. Imagine
that one starts with an initial state for S which is highly non classical, such
as a cat state $|{\rm cat}\rangle= 1/\sqrt{2} (| \alpha \rangle +
|- \alpha \rangle) $, where $|\alpha\rangle$ is a coherent state with large
amplitude ($|\alpha| \gg 1$). The corresponding Wigner function will have 
interference fringes, showing the quantum nature of the state. However, when
the system is put in contact with the environment and each  component
of the state becomes entangled with almost orthogonal states of the
environment, the
interference fringes will be washed out. 
In the end the Wigner function becomes positive defined, with two
peaks at the values corresponding to $\pm \alpha,$ as a true 
probability distribution. Decoherence transforms the initial
pure state into a mixture of the two coherent states $|\pm \alpha\rangle.$

Until not very long ago the ideas of decoherence were restricted to the 
theoretical domain. Recent experimental developments have succeeded in
studying in real time the process of decoherence in the laboratory, 
and have tested 
the predictions of the theory. Here we shall mention a few experiments that
have been a hallmark in those developments. First, in the field of cavity
QED, superposition states of photons and Rydberg atoms have been created within
high-Q microwave cavities. 
Cat states of around 3 photons have also
been produced, and it has been studied how they decay due to decoherence
\cite{haroche}. The coherence of the state was monitored 
with the help of a measurement of correlations between two consecutive
atoms crossing the cavity~\cite{davidovich}.
Second, in the field of ion trapping, methods for creating superposed 
motional states of ions were developed, as well as schemes of environment
engineering to protect those states from decoherence
\cite{wineland}. Finally,
it has been possible to push the size of the cat states further into the
macroscopic realm by generating a mesoscopic cat inside a rf-SQUID. The two
components of the cat correspond to  superconducting currents moving either
clockwise or counterclockwise, each containing around $10^9$ Cooper pairs
\cite{freeman}.

In what we have discussed so far decoherence has been portraited as a
``good'' effect, in the sense that is responsible for the quantum-classical
transition and the appearance of our classical world. Decoherence can also
have a ``bad'' role in the field of quantum computation and quantum information
processing. There one performs logical operations making use of the
superposition states of quantum mechanics. For such operations to be 
successful it is very important to maintain the relative phase between
the components of the superpositions all along the operations. 
If decoherence acts, it produces quantum
errors that must be somehow corrected. Several methods 
have been proposed to minimize the effects of decoherence (see~\cite{nmr} 
for an example in
nuclear magnetic resonance). 

The prototype calculation of environment induced decoherence is the
heuristic position-position interaction Hamiltonian for 
describing quantum Brownian motion, where the environment is 
taken to be a collection of harmonic oscillators. Although such
a model is quite useful for studying many  physical processes
associated to dissipation and decoherence of a quantum system,
the results that follow from it do not apply to every situation. 
That is, it is necessary to perform a case by case analysis in order to 
compute physical observables, such as decoherence and damping rates, how they scale with
the parameters of the system, the environment, and their coupling, etc. 
For usual environments (thermal atoms, thermal light, phonons, etc.) it is
in principle possible to design engineering schemes to protect the state of
the system from decoherence, for example by reducing the coupling to the
environment.

Then, the following question naturally arises: is it possible, at
least in principle, to have arbitrarily weak decoherence? In this paper, 
we consider a fundamental source of decoherence that cannot be 
`turned off':  the radiation pressure coupling with the vacuum field~\cite{rp2}.
As reviewed in Sec.~2, any particle not completely transparent 
unavoidably scatters vacuum field fluctuations. This type of coupling 
is responsible for the Casimir effect. More generally, photons are
created out of the vacuum field when moving boundaries are
considered, an effect known as dynamical Casimir effect or motion-induced 
radiation.
In Sec.~3, we show how the dynamical Casimir effect engenders decoherence. Our emphasis is on the 
basic physical ideas, and most of the 
calculations are referred to~\cite{rp1}, but we also briefly discuss a 
model  alternative to the one employed in this reference.

\section{Dynamical Casimir effect}

The Casimir effect is perhaps the simplest and most striking effect of the 
quantum vacuum field (see ~\cite{casimir-review} for reviews). 
The essential idea is that the boundary conditions 
modify the spectrum of the radiation field, and thereby its zero-point energy.
This modification has direct physical consequences, leading, for example, to 
an attractive force between two parallel perfectly-reflecting plates (of 
surfaces $A$) and at a distance $L,$ given by~\cite{casimir}
\[
F = {\pi^2 \over 240}  {\hbar c\over L^4}  A.
\]
A series of recent experiments~\cite{exp} reported precision measurements of 
the Casimir force in agreement with the 
predictions of Quantum Electrodynamics, although  more complete theoretical 
calculations, taking into account 
corrections due to finite temperature and conductivity as well as to  
roughness and geometry 
of the surfaces are partially yet to be done~\cite{theory-casimir}.     

The Casimir force may also be computed by taking the average of the 
Maxwell stress tensor over the field vacuum state~\cite{brown-maclay}.
 This method suggests that 
the Casimir force is itself a fluctuating quantity, as noted by Barton. Its 
fluctuations were first computed
for plane perfectly reflecting mirrors~\cite{Barton1}, and later for spherical
and spheroidal particles~\cite{Eberlein1}.
More generally, any particle scattering the radiation field is under the 
action of a fluctuating radiation pressure
force exerted by the vacuum field, even in the situations where the 
{\it average} force vanishes (for example a 
single plane mirror at rest).  The coupling responsible for those 
fluctuations also gives rise to a {\it dissipative} force, when the particle is
moving in vacuum. Dissipation of the 
mirror's mechanical energy is needed to enforce energy conservation, since the
motion induces the emission
of pairs of photons (for reviews see \cite{reviewdinamico}\cite{reviewdinamico2}). 
Because of their common physical origin, fluctuations and dissipation are 
related by a very general 
result~\cite{kubo}, whose most known application is the Einstein relation 
between diffusion and 
friction coefficients for a Brownian particle in the high-temperature limit. 
This connection provides a very useful tool for deriving the response 
 to an external small perturbation from the fluctuations in the 
unperturbed case. 
Linear response theory was employed by Jaekel and Reynaud to infer the 
vacuum radiation pressure force on  partially-reflecting 
moving mirrors  in the one-dimensional (1D) case~\cite{jaekel1}. 
For a single perfect mirror (position $x(t)$) the force is given by
\begin{equation}
F = {\hbar \over 6 \pi c^2} {d^3 x\over dt^3},\label{dinamico1}
\end{equation}
a result first obtained by solving the 
boundary conditions of a moving mirror in the long wavelength approximation, 
and assuming the effect of the motion to be  a small perturbation~\cite{ford}. 
Eq.~(\ref{dinamico1})  was also derived as the $n\rightarrow\infty$ limit
of a moving half-space of refractive index $n$~\cite{barton-eberlein}.
It also corresponds to the nonrelativistic approximation of the 
exact result (for a perfect mirror)
derived with the help of a conformal coordinate transformation to the 
co-moving frame~\cite{fulling}. 

Since the wave equation in three dimensions is not 
invariant under a general conformal transformation, only 
approximated methods are used in this case. 
The dissipative force on a plane mirror 
was computed within the long wavelength approximation
for a scalar~\cite{ford} and  electromagnetic~\cite{mn1} 
field models. The angular and frequency distributions of the 
emitted radiation were also computed for a 
single plane moving mirror~\cite{lasm}, a moving 
dielectric half-space~\cite{barton-north}~\cite{eberlein-die} and two 
parallel plane mirrors~\cite{mundarain}.
Linear response 
theory was employed to derive the dissipative force on moving 
spheres~\cite{mnr}. Small but otherwise arbitrary time-dependent deformations 
of an initially plane surface were analyzed with the help of different 
approaches: linear response theory~\cite{barton-p}, long wavelength 
approximation~\cite{bjp}, and path integrals~\cite{kardar}. 

The magnitude of the dynamical Casimir effect may be illustrated with
the following example, which we shall discuss in detail in Sec. 3. We consider
that
the `mirror' is a particle of mass $M$ in a 1D harmonic 
potential, such that
the oscillation frequency is $\omega_0.$ From Eq.~(\ref{dinamico1}), the 
equation of motion reads
\begin{equation}
{d^2 x\over dt^2} = - \omega_0^2 x + {\hbar \over 6 \pi M c^2} 
{d^3 x\over dt^3} .
\label{eq_motion}
\end{equation}    
For any situation of physical interest, the zero point energy is much smaller 
than the 
rest mass energy: $\hbar \omega_0\ll M c^2.$ In this case, (\ref{eq_motion})
has solutions  corresponding to oscillations damped at the rate 
\begin{equation}
\Gamma = {\hbar \omega_0\over 12 \pi M c^2} \omega_0 \ll \omega_0, 
\label{Gamma}
\end{equation}
showing that the dynamical Casimir effect provides a tiny perturbation of the 
free oscillations. 

As could be expected, a larger effect takes place when field modes of a cavity
resonator
are coupled to the moving boundaries, mainly when the mechanical 
frequency lies close to a given cavity eigenfrequency. Moore considered 
a scalar 1D field inside a cavity where one of the mirrors 
follows a 
prescribed motion~\cite{moore}. The field modes were formally built in terms 
of the 
solution of a functional equation. This method was later 
developed~\cite{cavity1} and 
extended to the case where the two mirrors are set in motion~\cite{dalvit}. 
The case of partially-transmitting mirrors was also calculated, allowing for 
a reliable estimation of the orders of magnitude for the rate of transmitted 
photons 
and the number of photons inside the cavity at steady-state~\cite{astrid}.
So far, few three-dimensional (3D) calculations along these lines have been 
reported. A rectangular cavity made of perfectly-reflecting moving 
mirrors~\cite{dodonov-3d} \cite{crocce},
and a spherical bubble with time-dependent radius~\cite{sono}   
were analyzed, the latter motivated by the problem of sonoluminescence. 

In this article we only consider a single scatterer, so that no resonant 
enhancement takes place. 
In this section, we have shown that the radiation pressure coupling 
gives rise to energy damping of a particle scattering vacuum fluctuations. In the
next section, we show that it also destroys the quantum coherence of the particle.

\section{Decoherence and the Casimir effect}

Most treatments of the dynamical Casimir effect consider the 
particle that scatters the vacuum field (the `mirror') 
to follow a prescribed motion (an exception is 
Ref.~\cite{fluct}, which considers 
fluctuations of position of a particle driven by  
vacuum radiation pressure). In this article, however,
we want to focus on the particle as the dynamical degree of freedom of interest. 
More specifically, we analyze how the radiation pressure coupling 
destroys the quantum coherence of an initial superposition state of the 
particle. 

We consider as before that the particle is in a harmonic potential well, corresponding to a frequency of
oscillation $\omega_0.$ The connection with the previous approaches, where the (classical) particle 
is assumed to follow a prescribed oscillation, is made by taking a coherent quasi-classical state 
$|\alpha\rangle$ for the particle, so that the combined particle-field state at $t=0$ is 
\begin{equation}
|\Psi(t=0)\rangle = |\alpha\rangle |0\rangle,
\end{equation} 
where we have assumed that the field is initially in the vacuum state $|0\rangle$.
The oscillation gives rise to the emission of photon pairs at time $t$ 
at the field modes $\lambda_1$ and 
$\lambda_2,$ with probability amplitudes $b(\lambda_1,\lambda_2,t):$
\begin{equation}    
|\Psi(t)\rangle = |\alpha\rangle \left(B(t)|0\rangle + \sum_{\lambda_1,\lambda_2}b(\lambda_1,\lambda_2,t)
|\lambda_1,\lambda_2\rangle\right),\label{psi1}
\end{equation} 
where $B(t)$ is such that this state is normalized. 
As discussed in Sec.~2 (see, in particular, Eq.~(\ref{Gamma})), 
the energy damping associated to the dynamical Casimir effect is  very small. This effect, and more generally 
the recoil of the particle, is  
neglected in~(\ref{psi1}), where the particle state is assumed not to be modified.
Even at this level of approximation, 
there is decoherence, as we show by taking the initial state of the 
particle to be the cat state $|{\rm cat}\rangle= (|\alpha\rangle + |-\alpha\rangle)/\sqrt{2},$
an example already mentioned in Sec.~1.
It corresponds to the coherent superposition 
of two wavepackets oscillating out-of-phase in the harmonic potential well.
The 
amplitudes  $b(\lambda_1,\lambda_2,t)$ depend on the phase of the oscillation, so that
they have an opposite sign when we take the state $|-\alpha\rangle.$ 
Since the evolution operator is linear, the complete state at time $t$ is the superposition of 
the r.-h.-s. of (\ref{psi1}) with the analogous state for $|-\alpha\rangle.$
It turns out to be 
 an entangled state of the 
form discussed in~(\ref{entangled}):
\begin{equation}
| \Psi(t) \rangle = |\alpha\rangle |\epsilon^{(+)}(t)\rangle +
                    |-\alpha\rangle |\epsilon^{(-)}(t)\rangle, \label{entangled2}
\end{equation}
with $|\epsilon^{(\pm)}(t)\rangle = B(t)|0\rangle \pm \sum_{\lambda_1,\lambda_2}b(\lambda_1,\lambda_2,t)
|\lambda_1,\lambda_2\rangle.$ These field states work as tags for the particle states, providing
which-way information about the phase of the oscillation. As time goes on, the information gets better
defined, since 
\begin{equation}
\langle \epsilon^{(-)}(t) | \epsilon^{(+)}(t)\rangle = 1 - 
2 \sum_{\lambda_1,\lambda_2}|b(\lambda_1,\lambda_2,t)|^2
\end{equation}
decreases as the probability for photon emission increases. 
When the emitted photons are not detected, all the relevant information about the particle is
contained in the reduced matrix $\rho(t)={\rm Tr}_{\rm F}(|\Psi(t)\rangle
\langle \Psi(t)|),$ where the trace is taken over the field states. 
Since the interference term is gradually washed out as a consequence of the photon emission effect
and the corresponding entanglement with the field, $\rho(t)$ decays into the 
statistical mixture $\rho_m = (|\alpha\rangle \langle\alpha| + |-\alpha\rangle \langle-\alpha|)/2.$
The corresponding time scale $t_d$ may be computed~\cite{rp1} from Eq.~(\ref{entangled2}), and turns
out to be proportional to the energy damping time $1/\Gamma,$ 
which is related to the two-photon probabilities by energy conservation:
\begin{equation}
t_d= {1\over 4 |\alpha|^2} {1\over \Gamma}.\label{main1}
\end{equation}
Eq.~(\ref{main1}) also holds when 
the coupling with the environment is  described by a heuristic master equation
in the Lindblad form (derived with the help of the rotating-wave approximation)~\cite{walls}, 
as well as
in the case of position-position coupling to a zero-temperature 
environment of harmonic oscillators, and has a very simple interpretation~\cite{caldeira}:
if $1/\Gamma$ is the time needed to damp the  energy $2|\alpha|^2 \hbar \omega_0,$  
it corresponds to the emission of $2|\alpha|^2$ pairs of photons (each pair 
has a total energy equal to $\hbar \omega_0$). On the other hand, coherence is
much more delicate than energy, since a single photon provides which-way information that
destroys the quantum phase of the cat state. Hence the decoherence time
is the time scale for the emission of a 
single photon. Since 
$4|\alpha|^2$ photons are emitted during the  time interval $1/\Gamma,$ the time 
for a single photon scales as in the r.-h.-s. of~(\ref{main1}).

Eq.~(\ref{main1}) only holds when $|\alpha|\gg 1.$ 
In this limit, decoherence is much faster than damping, justifying the 
approach of neglecting the decay of the amplitude $\alpha$ of the 
coherent states in (\ref{psi1}) and (\ref{entangled2}). 
This is of course in line with the idea that  in the `macroscopic' limit weird 
non-classical states are extremely fragile and difficult to observe. 
For truly macroscopic systems $t_d$ is so short that no
experimental monitoring of the decoherence process is possible. 
However the validity of Eq.~(\ref{main1}) is restricted by the additional
condition that decoherence is slower than the free oscillation 
(this condition is fulfilled by the experiments  \cite{haroche} \cite{wineland} discussed in Sec.~1).
In this regime, the particle oscillates several times in the potential well before 
coherence is lost, and the r.-h.-s. of Eq.~(\ref{main1}) may be written 
in terms of the distance $\Delta x = 2 \sqrt{2\hbar/M\omega_0} |\alpha| $ between the two 
wavepackets when they are at their turning points ($M$ is the mass of the 
particle): 
\begin{equation}
t_d= 4 \left( \Delta x_0\over \Delta x\right)^2 {1\over \Gamma},\label{main2}
\end{equation}
where $\Delta x_0=\sqrt{\hbar/(2 M \omega_0)}$ is the 
position uncertainty of the oscillator ground state. 
Eq.~(\ref{main2}) shows more explicitly that the decoherence rate scales as the 
squared distance in phase space between the two components of the cat state.
In Eq.~(\ref{main1}), the distance is expressed in terms of the squared difference 
$\Delta \alpha = 2\alpha$ between
the amplitudes of the two coherent states  $|\pm \alpha\rangle.$
Such dependence, already experimentally observed in~\cite{haroche}, was fully
verified in~\cite{wineland}. 
Thus, the decoherence rate 
is directly connected to the quality of which-way information, for the possibility of resolving the 
two wavepackets is quantified by the distance between them divided by
their width $\Delta x_0.$

The second factor entering in the r.-h.-s. of~(\ref{main1}) is the damping coefficient $\Gamma.$
Rather than a phenomenological constant, here $\Gamma$ quantifies 
the strength of the radiation pressure coupling to the vacuum field, and is calculated from first principles. 
As discussed in Sec.~2, it may be obtained directly from the expression for the dissipative 
radiation pressure force on the particle. In the 1D case, $\Gamma$ is given by
Eq.~(\ref{Gamma}), which jointly with Eq.~(\ref{main1}) yields
\begin{equation}
t_d = {3\over (v/c)^2} {2\pi\over \omega_0},\label{main-casimir}
\end{equation}
where $v = \sqrt{2\hbar \omega_0/M}\, |\alpha|$ is the velocity of the wavepackets at the moment they
cross the bottom of the potential well. 
Therefore, in the nonrelativistic limit considered in this paper, 
decoherence is much slower than the free oscillation. 
The ratio between the two time
scales is even larger when considering the real 3D case. 
If we take a spherical perfectly-reflecting particle of radius $R$ 
smaller than the range of oscillation, then $\omega_0 R/c < v/c \ll 1.$
Since the relevant field modes have frequencies of the order of
$\omega_0,$ in this limit the particle
is much smaller than typical wavelengths (Rayleigh scattering regime), and 
hence is weakly coupled to the field.
The dissipative force in this regime was calculated in~\cite{mnr}; the 
resulting damping coefficient scales as the squared polarizability of the 
sphere, leading to an additional factor $(\omega_0 R/c)^6:$  
\begin{equation}
t_d = {324\over (v/c)^2} \left( c \over \omega_0 R \right)^{6} 
{2\pi\over \omega_0}\gg \left(\frac{c}{v}\right)^8 
{2\pi\over \omega_0} . \label{vacuo}
\end{equation}

It is also possible to analyze the decoherence effect in a more complete theoretical
framework, where the dynamical radiation pressure coupling between particle and field
is fully taken into account. This approach also accounts properly for damping of the
particle's energy, as well as for additional effects resulting from the coupling with the 
field.  Moreover, it allows us to analyze decoherence in the more general case of an 
arbitrary temperature of the field. 
An ab-initio Hamiltonian model for the particle-field system 
was derived from first principles in Ref.~\cite{barton}. This model was the
starting point for the discussion of decoherence in Refs.~\cite{rp2} and \cite{rp1}. 
The field scattering corresponds to frequency dependent reflection and transmission coefficients 
that satisfy the passivity requirements discussed in~\cite{passivity}.
This means that the dynamics of the particle does not suffer from the instabilities 
associated to the model of a perfect mirror  (as well known from classical electron theory, 
Eq.~(\ref{eq_motion}) is plagued with `runaway' solutions).

Here we describe the radiation pressure coupling with the alternative, more intuitive model, where
the interaction Hamiltonian corresponds to the energy transfer between field and particle:
\begin{equation}
H_{\rm int}= - x F , \label{model}
\end{equation}
where $F$ is the radiation pressure force on the particle, and $x$ its position.
This type of model was extensively employed in several contexts associated to the dynamical
Casimir effect~\cite{reviewdinamico}. 
Here we focus on the limit where the particle perfectly reflects the (1D)
field, but a discussion of partially-reflecting mirrors along these lines is also
possible. As shown below, it leads to results for the decoherence and damping rates in 
agreement with those found in Ref.~\cite{rp1}.
  
Starting from (\ref{model}), we derive a master equation for the reduced density matrix 
of the particle. It is similar to the master equation for QBM, derived from the position-position 
interaction Hamiltonian. Technically, the essential difference arises from the fact that the 
force operator $F$ in (\ref{model}) is quadratic in the field operators, which leads to a
damping coefficient that depends on the state (and hence temperature) of the field (reservoir). 
Although the formalism relies on a 1D model, the final results may be 
generalized to the 3D case. 

We write the master equation 
in terms of the Wigner function $W(x,p,t):$
\begin{equation}
\partial_t W=-{p\over M} \partial_xW + M\omega_0^*{}^2 x \partial_pW +2\Gamma\partial_p(pW)+D_1{\partial^2W\over \partial p^2}
-D_2{\partial^2W\over \partial x \partial p}.
\label{FP}
\end{equation}
The first two terms in (\ref{FP}) correspond to the harmonic oscillation in the potential well, with a 
frequency $\omega_0^*=\omega_0+\delta\omega$ 
modified by the coupling with the field  (on the other hand, 
when the interaction Hamiltonian is linear in the momentum of the particle, a mass correction appears~\cite{barton}).
The remaining terms describe non-unitary evolution. The damping as well as the diffusion coefficients $D_1$ and $D_2$
are time dependent and 
given in terms of correlation functions of the force operator. The diffusion coefficients are related to the
symmetric correlation function:
\begin{equation}
\sigma_{FF}(t)=\langle\{F(t),F(0)\}\rangle ,
\end{equation}
where the brackets denote the anticommutator, and the average is taken over the field state (thermal 
equilibrium, temperature $T$). 

The term proportional to $D_2$ in~(\ref{FP}) yields a negligible contribution, so that we 
focus on $D_1:$
\begin{equation}
D_1(t)= {1\over 2}\int {d\omega\over 2\pi} \sigma_{FF}[\omega]\,{\rm sync}_t(\omega), \label{D13} 
\end{equation}
where $\sigma_{FF}[\omega]$ is the Fourier transform of $\sigma_{FF}(t)$
and $${\rm sync}_t(\omega)={\sin[(\omega-\omega_0)t]\over \omega-\omega_0}$$
is a function peaked around $\omega=\omega_0$ of width $2\pi/t.$
Clearly, for a time $t$ long enough, the function ${\rm sync}_t(\omega)$ is so sharply
peaked that $\sigma_{FF}[\omega]$ is approximately constant over the short frequency interval that
contributes in the integral in Eq.~(\ref{D13}), and hence may be 
replaced by its value at $\omega=\omega_0.$  In this case, we find
\begin{equation}
D_1(t\rightarrow \infty)= {1\over 4} \sigma_{FF}[\omega_0]. \label{D12} 
\end{equation}    
A sufficient (and also necessary at $T=0$) 
condition for the validity of (\ref{D12}) is $\omega_0 t \gg 1.$ In other words, 
for times much longer than the period of oscillation, the field fluctuations at frequency 
$\omega_0$ provide the dominant contribution to diffusion. 

The damping coefficient is likewise connected to the average value of the {\it commutator} of the force operator
taken at different times (anti-symmetric correlation function). 
When the interaction Hamiltonian is linear in the operators of the environment, as in the 
position-position model, the commutator is a c-number times a delta function (in time), and as a consequence, 
the damping coefficient has a constant value
that does not depend on the state of the environment. As already mentioned, 
this is not the case for radiation pressure coupling. In particular, the damping coefficient depends on the 
temperature of the field, as could be expected having in mind the Stefan-Boltzmann law. 
At zero temperature, we recover the result given by Eq.~(\ref{Gamma}).

We calculate the pointer states  using the predictability
sieve criterion, discarding all information about the environment, as discussed in Sec.~1.
We start from the master equation, and evaluate the rate of change of linear entropy, assuming an
initial pure state. 
It is straightforward to show that the entropy is minimized 
for minimum uncertainty Gaussian states,
hence the pointer states are the coherent states. 
This result agrees with the well-known fact that coherent states provide
the closest possible realization of a classical state of oscillation, 
given the constraint imposed by the Heisenberg uncertainty
relation. In short, coherent states remain approximately pure because they do not entangle with 
field states, at least for times shorter than the damping time $1/\Gamma,$ as shown by Eq.~(\ref{psi1}).

On the opposite extreme in Hilbert space, superpositions of coherent states are
highly nonclassical and cannot last when the distance between the two
components is large. 
This may be analyzed in detail from Eq.~(\ref{FP}). 
The coherence of the initial  state  is imprinted on the 
Wigner function  in the form of an interference term $W_{\rm int}$
that oscillates in
phase space. When the two state components  are spatially separated by a 
distance $\Delta x,$ the oscillation is along the  axis of momentum:
$W_{\rm int}(x,p)\sim \cos(\Delta x \,p/\hbar).$
Thus, according to Eq.~(\ref{FP}), 
diffusion washes out this oscillatory term, the faster 
the larger the value of $\Delta x.$  With an additional 
factor of $2$ to take into account the average over several
free rotations of the state in phase space, we find
\begin{equation}
t_d = 2 {\hbar^2\over D_1 (\Delta x)^2}. \label{deco}
\end{equation}
To derive the decoherence time when the field is in the vacuum sate,  
we compute the correlation function $\sigma_{FF}[\omega_0]$ at zero 
temperature. When replacing the result for $D_1$ as given by (\ref{D12}) into
(\ref{deco}), 
we obtain the same result already derived in this Section by a more elementary method.

For finite temperatures, 
the spectrum  is approximately constant at low frequencies, so that 
(\ref{D12}) also holds 
when $\omega_0\ll k T/\hbar$
($k$ is the Boltzmann
constant), including the 
free particle limit $\omega_0=0,$ 
provided that the entire frequency interval  around $\omega_0$
is contained in the low frequency part of the spectrum, which
corresponds to the condition $1/t \ll k T/\hbar.$ 
The damping coefficient may calculated in the high temperature limit as well, and
the results are in agreement with Einstein relation 
\begin{equation}
D_1 = 2 M k T \Gamma.\label{Einstein}
\end{equation}          
More generally, we may derive a relation between diffusion and damping coefficients 
valid for arbitrary values of temperature~\cite{rp1}, including $T=0,$ starting from 
the general relation between symmetric and anti-symmetric correlation functions 
(fluctuation-dissipation
theorem).
 
The decoherence time for high $T$ is derived by
replacing (\ref{Einstein}) into (\ref{deco}). As we discuss below, usually in this 
limit  decoherence is faster than the free oscillation, so that, contrary to
the $T=0$ case, there is no average over many oscillations in this case.
To describe the decoherence
process, we must evaluate the diffusion coefficient at a time $t$ much shorter
than $t_d.$ Hence, we are allowed to use its asymptotic value 
as given by the Einstein relation (\ref{Einstein}) only if we
assume that $t_d\gg \hbar/(k T).$
The resulting expression is  very general~\cite{WHZ1}, and also
holds in the free particle case:
\begin{equation}
t_d =  {\lambda_T^2\over (\Delta x)^2}{1\over \Gamma},\label{tdHT}
\end{equation}
where  $\lambda_T = \hbar/\sqrt{2 M k T}$ 
is the de Broglie wavelength of a particle of mass $M$ in thermal equilibrium. 
Eq.~(\ref{tdHT}) has a form similar to (\ref{main2}),
except that now the 
reference of distance is set by  thermal fluctuations instead of 
zero point fluctuations. 

In order to complete the evaluation of the decoherence time, 
we need to evaluate the damping coefficient $\Gamma$ in the 
high temperature limit.
We consider as before a sphere of radius $R,$ 
which is usually much larger than 
typical field wavelengths, which  are of the order of $\hbar c/(kT)$ 
(except for very low temperatures or very small spheres). In this
short-wavelength regime, the radiation pressure force 
may be calculated by replacing the surface of the sphere by
a collection of tangent planes, and the final result reads
\begin{equation}
F= -{4 \pi^3\over 45} {(kT)^4\over  \hbar^3 c^4} R^2 {dx\over dt}.\label{FHT}
\end{equation}
The force scales with the surface or cross section of the sphere, 
and is proportional to $T^4,$ in agreement with
Stefan-Boltzmann law. As opposed to the vacuum case,  here we have  a 
true friction force, i.e. proportional to the velocity of the particle 
and not to higher-order time derivative as in 
Eq.~(\ref{dinamico1})
(the thermal field is not Lorentz invariant).

In the free case ($\omega_0=0$), $\Gamma$ is simply the coefficient 
multiplying the velocity in Eq.~(\ref{FHT}) divided by $M.$ Then,
with the help of 
(\ref{tdHT}) we find
\begin{equation}
t_d = {45\over 8 \pi^3}{\hbar^5 c^4 \over (kT)^5 R^2 (\Delta x)^2}.\label{final}
\end{equation}
Eq.~(\ref{final}) shows that the decoherence time depends strongly on temperature
(the same temperature dependence was found in Ref.~\cite{joos}). 
Even at the temperature corresponding to the cosmic background radiation, $T=2.7 {\rm K},$   
radiation pressure is a very efficient source of decoherence. As an example, for 
$R= 1 {\rm cm},$ we have $t_d = 2.7\times 10^{-21}/(\Delta x[{\rm m}])^2 \, {\rm s},$
which is in the nanosecond range for a separation $\Delta x = 1 \mu{\rm m}.$

\section{Conclusion}

The master equation provides a complete description of the particle dynamics 
when no measurement on the field is made. It accounts for the renormalization of
the oscillation frequency, damping, and diffusion and the associated decoherence effect. It also
allows for the determination of the pointer states, and all that for any temperature $T.$ 
On the other hand, the decoherence time scale at $T=0$ may be calculated by a simpler approach, 
in which we follow the evolution of the complete particle-field state, 
calculated with the help of the superposition principle,
and trace over the field
at the very end. This approach explicitly shows that decoherence results from entanglement between 
particle and field states. 

The decoherence induced by radiation pressure coupling with vacuum 
fluctuations is a very slow effect, when compared with the 
the time scale of the free evolution. Yet, it is remarkable, 
from a conceptual point-of-view, 
that classical behavior of a macroscopic system emerges from the 
formalism of Quantum Mechanics itself, even though in very
long time scale, provided that 
the quantum vacuum radiation field is taken into account.

P.A.M.N. thanks M.T. Jaekel, A. Lambrecht and S. Reynaud 
for discussions, and CNPq, PRONEX  and FAPERJ for partial financial support.


\begin{thebibliography}{800.}
\addcontentsline{toc}{section}{References}

\bibitem{giulini} D. Giulini {\it et al}: \emph{Decoherence and the Appearance
of a Classical World in Quantum Theory} (Springer, Berlin 1996) 
H.D. Zeh: Found. Phys. {\bf 3}, 109 (1973)

\bibitem{WHZ1}  W.H. Zurek: Phys. Today {\bf 44}, No. 10, 36 (1991)

\bibitem{WHZ2} W.H. Zurek: Phys. Rev. D {\bf 24}, 1516 (1981)

\bibitem{WHZ3} W.H. Zurek: Prog. Theor. Phys. {\bf 89}, 281 (1993)

\bibitem{QBM}  W.G. Unruh and W.H. Zurek: Phys. Rev. D {\bf 40}, 1071
(1989) B.L. Hu, J.P. Paz and Y. Zhang: Phys. Rev. D {\bf 45}, 2843 (1992)

\bibitem{ZHP} W.H. Zurek, S. Habib, and J. P. Paz: Phys. Rev. Lett. {\bf 70},
1187 (1993)

\bibitem{ZP} W.H. Zurek and J. P. Paz: Phys. Rev. Lett. {\bf 82}, 5181 (1999)

\bibitem{CME} D.A.R. Dalvit, J. Dziarmaga and W.H. Zurek, to appear in
Phys. Rev. Lett.

\bibitem{haroche}  M. Brune {\it et al}: Phys. Rev. Lett. {\bf 77}, 4887
(1996) A. Rauschenbeutel {\it et al.}: Science {\bf 288}, 2024 (2000)

\bibitem{davidovich} L. Davidovich {\it et al}: Phys. Rev. A {\bf 53}, 1295 (1996)

\bibitem{wineland} C.J. Myatt {\it et al}: Nature {\bf 403}, 269 (2000)
C. A. Sackett {\it et al}: Nature {\bf 404}, 256 (2000) 

\bibitem{freeman} J. Friedman {\it et al}: Nature {\bf 406}, 43 (2000)
C.H. van der Wal {\it et al}: Science {\bf 290}, 773 (2000)

\bibitem{nmr} D.G. Cory {\it et al}: Phys. Rev. Lett. {\bf 81}, 2152 (1998)

\bibitem{rp2} 
 D.A.R. Dalvit and P.A. Maia Neto: Phys. Rev. Lett. {\bf 84},  798 (2000)

\bibitem{rp1} P.A. Maia Neto and  D.A.R. Dalvit: Phys. 
Rev. A {\bf 62}, 042103 (2000)

\bibitem{caldeira} A.O. Caldeira and A.J. Leggett: Phys. Rev. {\bf A} 31,
1059 (1985)

\bibitem{casimir-review} S.K. Lamoreaux: Am. J. Phys. {\bf 67}, 850 (1999)
G. Plunien, B. M\"uller and W. Greiner: Phys. Rep. {\bf 134}, 87 (1986)
V.M. Mostepanenko and N.N. Trunov: {\it The Casimir Effect 
and its Applications} (Clarendon, London, 1997)

\bibitem{casimir} H.B.G. Casimir: Proc. K. Ned. Akad. Wet. {\bf 51}, 793 (1948)

\bibitem{exp} S.K. Lamoreaux: Phys. Rev. Lett. {\bf 78}, 5 (1997)
U. Mohideen and A. Roy: Phys. Rev. Lett. {\bf 81}, 4529 (1998)

\bibitem{theory-casimir} A. Lambrecht and S. Reynaud: 
Eur. Phys. J. D {\bf 8}, 309 (2000)
C. Genet, A. Lambrecht and S. Reynaud: Phys. Rev. A {\bf 62}, 012110 (2000)
M. Bordag {\it et al}:  Phys. Rev. Lett. {\bf 85}, 503 (2000)

\bibitem{brown-maclay} L.S. Brown and G.J. Maclay: Phys. 
Rev. {\bf 184}, 1272 (1969)

\bibitem{Barton1} G. Barton: J. Phys. A: Math. Gen. {\bf 24}, 991 (1991) 
J. Phys. A: Math. Gen. {\bf 24}, 5533 (1991)

\bibitem{Eberlein1} C. Eberlein: J. Phys. A: Math. Gen. {\bf 25}, 3015 (1992) 
J. Phys. A: Math. Gen. {\bf 25}, 3039 (1992)

\bibitem{JR1} M.T. Jaekel and S. Reynaud: Quantum Opt. {\bf 4}, 39 (1992)


\bibitem{JR} M.T. Jaekel and S. Reynaud: Phys. Lett. A {\bf 172}, 319
(1993)

\bibitem{barton}  G. Barton and A. Calogeracos: Ann. Phys. (NY) {\bf 238},
227 (1995) A. Calogeracos and G. Barton: Ann. Phys. (NY) {\bf 238}, 268
(1995)

\bibitem{reviewdinamico} M.T. Jaekel and S. Reynaud: Rept. 
Prog. Phys. {\bf 60}, 863 (1997)

\bibitem{reviewdinamico2}
M. Kardar and R. Golestanian: Rev. Modern Phys. {\bf 71}, 1233 (1999)

\bibitem{kubo}  H.B. Callen and T.A. Welton:   Phys. Rev. {\bf 83},
34 (1951) R. Kubo: Rep. Progr. Phys. {\bf 29}, 255 (1966)

\bibitem{jaekel1} M.T. Jaekel and S. Reynaud: Quantum Opt. {\bf 4}, 39 (1992)

\bibitem{ford} L.H. Ford and A. Vilenkin: Phys. Rev. D {\bf 10}, 2569 (1982)

\bibitem{barton-eberlein} G. Barton and C. Eberlein: 
Ann. Phys. (N.Y.) {\bf 227}, 222 (1993)

\bibitem{fulling} S.A. Fulling and P. Davies: Proc. R. Soc. 
London Ser. A {\bf 348}, 393 (1976)

\bibitem{mn1} P.A. Maia Neto: J. Phys. A: Math. Gen. {\bf 27}, 2167 (1994)

\bibitem{lasm} L.A.S. Machado and P.A. Maia Neto: Phys. 
Rev. A {\bf 54}, 3420 (1996)

\bibitem{barton-north} G. Barton and C. North: Ann. Phys. (N.Y.) {\bf 252}, 
72 (1996)

\bibitem{eberlein-die} R. G\"{u}tig and C. Eberlein: J. Phys. A: Math. 
Gen. {\bf 31}, 6819 (1998)

\bibitem{mundarain} D.F. Mundarain and P.A. Maia Neto: Phys. Rev. A {\bf 57}, 
1379 (1998)

\bibitem{mnr} P.A. Maia Neto and S. Reynaud: Phys. Rev. A {\bf 47}, 1639 
(1993) 

\bibitem{barton-p} G. Barton: `New Aspects of the Casimir Effect'. In:
\emph{Cavity Quantum Electrodynamics, Supplement: Advances in Atomic, 
Molecular and Optical Physics}, ed. by P.R. Berman (Academic Press, 
New York 1993)

\bibitem{bjp} P.A. Maia Neto and L.A.S. Machado: 
Brazilian J. Phys. {\bf 25}, 324 (1995) 

\bibitem{kardar} R. Golestanian and M. Kardar: 
Phys. Rev. Lett. {\bf 78}, 3421 (1997)
R. Golestanian and M. Kardar: Phys. Rev. A {\bf 58}, 1713 (1998)

\bibitem{moore} G.T. Moore: J. Math. Phys. {\bf 11}, 2679 (1970)

\bibitem{cavity1} D.A.R. Dalvit and F.D. Mazzitelli: Phys. Rev. A {\bf 57}, 
2113 (1998) C.K. Cole and W.C. Schieve: Phys. Rev. A {\bf 52}, 4405 (1995)
V.V. Dodonov, A.B. Klimov and D.E. Nikonov: J. Math. Phys. {\bf 34}, 
2742 (1993).

\bibitem{dalvit}  D.A.R. Dalvit and F.D. Mazzitelli: Phys. 
Rev. A {\bf 59}, 3059 (1999)

\bibitem{astrid} A. Lambrecht, M.T. Jaekel and S. Reynaud: Phys. Rev. Lett. 
{\bf 77}, 615 (1996)

\bibitem{dodonov-3d} V.V. Dodonov and A.B. Klimov: Phys. Rev. A {\bf 53}, 
2664 (1996)

\bibitem{crocce} M. Crocce, D.A.R. Dalvit and F.D. Mazzitelli: {\it xxx archives}
quant-ph/0012040 (2000)

\bibitem{sono} C. Eberlein: Phys. Rev. A {\bf 53}, 2772 (1996) Phys. Rev. 
Lett. {\bf 76}, 3842 (1996)

\bibitem{fluct} M.T. Jaekel and S. Reynaud: J. Phys. France {\bf I 3}, 1
(1993)

\bibitem{walls} D.F. Walls and G.J. Milburn: Phys. Rev. A {\bf 31}, 2403 (1985)

\bibitem{passivity} M.T. Jaekel and S. Reynaud: Phys. Lett. A {\bf 167}, 
227 (1992)
A. Lambrecht, M.T. Jaekel and S. Reynaud: Phys. Lett. A {\bf 225}, 188 (1997)

\bibitem{2mirrors} M.T. Jaekel and S. Reynaud: J. Phys. France {\bf I 2}, 149
(1992)

\bibitem{joos} E. Joos and H.D. Zeh:  Z. Phys. B {\bf 59}, 223 (1985)


\end{thebibliography}
\end{document}